\documentclass[a4paper, 11pt]{article}
\pdfoutput=1 % if your are submitting a pdflatex (i.e. if you have
\usepackage{jheppub} % for details on the use of the package, please
\usepackage{color}

\newcommand{\be}{\begin{equation}}
\newcommand{\ee}{\end{equation}}
\def\beqa{\begin{eqnarray}}
\def\eeqa{\end{eqnarray}}
\def\nn{\nonumber}

\newcommand{\R}{\mathbb{R}}

\newcommand{\eqn}[1]{(\ref{#1})}
\newcommand{\del}{\partial}
\newcommand{\Tr}[1]{\:{\rm Tr}\,#1}
\newcommand{\dd}{{\mathrm d}}
\title
{\boldmath Noncommutative $\R^d$ via closed star product}
\author[a,b]{V.G. Kupriyanov and}
\author[c,d]{P. Vitale}
\affiliation[a]{Universidade Federal do ABC, Brasil\\
{\sl CMCC}, 09210-580 Santo Andre, SP, Brazil}
\affiliation[b]
{Tomsk State University, Tomsk 634050, Russia}
\affiliation[c]{Dipartimento di Fisica, Universit\`{a} di Napoli
{\sl Federico II}\\
Monte S.~Angelo, Via Cintia, 80126 Napoli, Italy}
\affiliation[d]
{INFN, Sezione di Napoli
\\
Monte S.~Angelo, Via Cintia, 80126 Napoli, Italy}

\emailAdd{vladislav.kupriyanov@gmail.com}
\emailAdd{patrizia.vitale@na.infn.it}

\abstract{
 We consider linear star products on $\R^d$ of Lie algebra type. First we derive the closed formula for the polydifferential representation of the corresponding Lie algebra generators. Using this representation we define the Weyl star product on the dual of the Lie algebra. Then we construct a gauge operator relating the Weyl  star product with the one which is closed with respect to some trace functional, $\Tr( f\star g)= \Tr( f\cdot g)$. We introduce the derivative operator on the algebra of the closed star product and show that the corresponding Leibniz rule holds true up to a total derivative. As a particular example we study the space $\R^3_\theta$ with $\mathfrak{su}(2)$ type noncommutativity and show that in this case the closed star product is the one obtained from the Duflo quantization map. As a result a Laplacian can be defined such that its commutative limit reproduces the ordinary commutative one. The deformed Leibniz rule is applied to scalar field theory  to derive conservation laws and the corresponding noncommutative currents.}

\begin{document}
\maketitle
\flushbottom

\section{Introduction}
There has  recently been a renewed interest in noncommutative structures on $\R^3$ of Lie algebra type (in particular $\mathfrak{su}(2)$),
in connection with their occurrence in three-dimensional quantum gravity models \cite{QGrav}, where $\R^3$
is  identified with
the dual algebra of the  local relativity group. In such a framework star products are mainly introduced through a group Fourier transform (see for example \cite{OR13} for a review and comparison with other techniques),  imposing compatibility with
the group convolution.

Besides their appearence in the quantum gravity context, noncommutative structures on
$\R^3$ are very interesting because of their application in the quantization of standard dynamical systems as for example the  hydrogen atom \cite{RV,GP13,Kup15}.  As for their implications in quantum field theory, such
as their renormalization properties and the UV/IR behaviour, which are quite different from
noncommutative field theories on Moyal space-time, they are analyzed in \cite{Vitale1,Vitale2}within the noncommutative space $\mathbb{R}^3_\lambda$ first introduced in \cite{Jabari}. The latter  may be regarded as a sequence of fuzzy spheres $\mathbb{R}^3_\lambda= \oplus_j \mathbb{S}_j$, with different radii labelled by $j\in \mathbb{N}/2$. The coordinate functions of such a noncommutative algebra  satisfy  $\mathfrak{su}(2)$  type star commutators
 \begin{equation}\label{1}
    \left[ {x}^{i},{x}^{j}\right]_{\star_\lambda}=i \lambda \varepsilon ^{ijk}{x}^{k}
\end{equation}
where the star product is the one in  \cite{Jabari}; it  turns out not to be closed with respect to any natural integration measure  \cite{Jabari,Vitale1,Vitale2}, that is $\Tr(f\star_\lambda g)\neq \Tr(fg)$, though being cyclic.  Derivations in this algebra, are inner,  $\hat D_i=[x_i,\;\cdot\;]_{\star_\lambda}$. The definition of a Laplacian in terms of such derivations  does not reproduce the ordinary Laplacian on $\mathbb{R}^3$, whereas the fact that the star product is not closed complicates the calculation of  free dynamics already for a simple scalar field theory (see \cite{Vitale1} for details).  To overcome the latter  a   matrix basis adapted to $\mathbb{R}^3_\lambda$ has been proposed in \cite{Vitale1}. Then the constructed framework was applied \cite{Vitale1} to study one loop contributions to the two point correlation function of real valued scalar noncommutative field theories with quartic polynomial interaction. No singularity of $IR$ type was found, which signals the absence of $UV/IR$ mixing in the considered theory.  In \cite{Vitale2} the  formalism has been  extended to include $U(1)$  gauge  theory on such a space at one loop.

One of the  questions which has motivated the present work   is  whether  the absence of mixing found in  \cite{Vitale1} is a consequence of the specific choice of the deformed Laplacian, or it is a general property of  NC field theories on deformed $\R^3$ with star products reproducing the star commutator \eqn{1} (that is, star products of $\mathfrak{su}(2)$ type). To this, we shall look for yet another $\mathfrak{su}(2)$ based star product, with the notable property of being closed with respect to the corresponding trace functional,
 \begin{equation}\label{2}
    \Tr\left( f\star g\right)= \Tr\left( f\cdot g\right).
\end{equation}
This property will make it possible to define the Laplacian, hence the kinetic part of any classical action of fields on noncommutative  $\R^3$, just as in the commutative case, $\Delta=\partial^2_x+\partial^2_y+\partial^2_z$. This will settle the issue of the commutative limit which was pointed out in \cite{Vitale1,Vitale2}.
The new star product,  $\star$,
is equivalent to $\star_\lambda$ in the sense that they agree at the level of star commutator of coordinates, Eq. \eqn{1} and  an invertible map  relates the two.
We will  indicate with $\R^3_\theta$ the noncommutative algebra with the new star product,  $(\mathcal{F}(\R^3),\star)$, to distinguish it from the one in \cite{Jabari, Vitale1,Vitale2}.

The existence of a star product closed with respect to the corresponding trace functional, which is known as the generalized Connes-Flato-Sternheimer conjecture, is proven in  \cite{Felder}.\footnote{ For convenience we report the result

\noindent{\bf Theorem \cite{Felder}} Let $\mathcal{M}$ be a Poisson manifold with the bivector field $\omega^{ij}\left( x\right)$, and let ${\bf\Omega}$ be
any volume form on $\mathcal{M}$ such that $\mathrm{div}_\Omega \omega=0$. Then there exists a star product on
$C^\infty(\mathcal{M})$ such that for any two functions $f$ and $g$ with compact support one has:
\begin{equation}\label{FS}
    \int (f\star g)\cdot {\bf\Omega}=   \int  f\cdot g\cdot{\bf\Omega}.
\end{equation}}
The problem is how to find it. The idea, see e.g. \cite{Kup15}, is to start with some appropriate star product (here the Weyl ordered one)  and then use the gauge freedom \cite{Kontsevich} in the definition of the star product to obtain the desirable one. Indeed if $\star$ and $\star'$ are two different star products corresponding to the same Poisson bi-vector $\omega^{ij}\left( x\right)$, they are related by a local  transformation
\begin{equation}
T\left( f\star g\right)=\left( Tf\star' Tg\right) ,  \label{gauge}
\end{equation}
where $T=1+O(\theta)$ is what we shall call the gauge operator. An instance of such a procedure can be found in \cite{Dito}, where  a gauge operator $T$ was constructed,  realizing the equivalence between the Gutt star product \cite{Gutt} and the Kontsevich one on the dual of  Lie algebras.

Our first goal is to construct  the Weyl star product $\star_W$, for a linear Poisson structure. To this end in Section 2 we derive the closed formula for the polydifferential representation of the algebra (\ref{1}), Eq. \eqn{31}.

In Section 3 we generalize the proposed construction to the case of $d$ dimensional Lie algebras,  $\left[ \hat{x}^{i},\hat{x}^{j}\right] =i \theta f ^{ij}_k\hat{x}^{k}$. Then, in Section 4, using the proposed polydifferential representation, we define the Weyl star product on the dual of  Lie algebras. We compare our results with the existing ones known in the literature, in particular with the one obtained from the Baker-Campbel-Hausdorff (BCH) formula.

The general procedure of the construction of the gauge operator $T$ which transforms a given star product $\star ^{\prime}$ in the closed one $\star$ is proposed in  Section 5. Thus  in Section \ref{gaugeweyl} we apply it to the particular example of the $\mathfrak{su}(2)$ algebra.  The gauge operator relating the Weyl star product with the closed one is given by Eq. \eqn{51}. We prove that it coincides with the Duflo correction to the Poincar\'e-Birkhoff-Witt map.
We thus derive  the explicit formula for the closed star product on $\R^3_\theta$ and show that it is exactly the star product obtained from the Duflo quantization map.
In section \ref{derop} we introduce the derivative operators $\hat\partial_i$ on the algebra of functions with closed star product and discuss their  properties. In  general, the Leibniz rule will be violated, but in a controlled way \cite{Kup17}.

To conclude  we shortly consider in Section 8 an application of our formalism to scalar field theory on $\mathbb{R}^3_\theta$  and outline the study its classical properties.

\section{Polydifferential representation for $\mathfrak{su}(2)$}

To start with, let us construct the polydifferential representation for the algebra of operators
 \begin{equation}\label{su2}
    \left[ \hat{x}^{i},\hat{x}^{j}\right] =i \theta \varepsilon ^{ijk}\hat{x}^{k}.
 \end{equation}
 The perturbative construction of the representation of the noncommutative algebra corresponding to any Poisson bi-vector $\omega^{ij}(x)$, suitable for the definition of the Weyl star product was proposed in \cite{KV}. In \cite{Gutt,Meljanac}  a universal formula was given for the  representation of Lie algebra generators as formal power series of the corresponding structure constants with the coefficients in Bernoulli numbers. In \cite{Meljanac1} the above procedure was applied to obtain the closed formula of the polydifferential representation for the $\kappa$-Minkowski Lie algebra on $\R^d$. Here we will follow the same approach. We will look for a representation of the algebra (\ref{su2}) in the form
\begin{equation}\label{22}
    \hat x^i=x^lg_{il}(\partial),
\end{equation}
where the functions $g_{il}(\partial)$, satisfy the equation
\begin{equation}\label{23}
   g_{jl}\partial^lg_{ik}-g_{il}\partial^lg_{jk}=i\theta\varepsilon^{ijl}g_{lk},
\end{equation}
with $\partial^l=\delta/\delta(\partial_l)$.

Perturbative calculations of the solution of (\ref{23}) suggest the following anzatz
\begin{equation}\label{24}
  g_{il}(\partial)=\delta_{il}-\theta \phi_{il}(\partial),
\end{equation}
where
\begin{equation}\label{25}
  \phi_{il}(p)=\frac{i}{2}\varepsilon^{ilm}p_m+\frac{\theta}{12}\left(\delta^{il}p^2-p_ip_l\right)\chi\left(\frac{\theta^2p^2}{2}\right).
\end{equation}

Substituting (\ref{24}) in (\ref{23}) we obtain the equation
\begin{equation}\label{26}
   \partial^i \phi_{jk}- \partial^j \phi_{ik}+\theta(\phi_{jl}\partial^l\phi_{ik}-\phi_{il}\partial^l\phi_{jk})=i\varepsilon^{ijk}-i\theta\varepsilon^{ijl}\phi_{lk}.
\end{equation}
On using (\ref{25}) we compute
\begin{equation}\label{27}
 \partial^i \phi_{jk}- \partial^j \phi_{ik}= i\varepsilon^{ijk}+\frac{\theta}{12}\left(p_i\delta^{jk}-p_j\delta^{ik}\right)(3\chi+\theta^2p^2\chi'),
\end{equation}
where $\chi'$ indicates the ordinary derivative of the function $\chi$ with respect to the argument. Also we calculate
\begin{align}\label{28}
 &\phi_{jl}\partial^l\phi_{ik}-\phi_{il}\partial^l\phi_{jk}\\
 &= - \left(p_i\delta^{jk}-p_j\delta^{ik}\right)\left(\frac{1}{4}+\frac{\theta^2\chi^2p^2}{144}\right)-\frac{i\theta \chi}{12}\varepsilon^{ijk}p^2+\frac{i\theta \chi}{12}\varepsilon^{ijl}p_lp_k.\notag
\end{align}
The RHS  of (\ref{26}) can be written as
\begin{equation}\label{29}
  i\varepsilon^{ijk}-i\theta\varepsilon^{ijl}\phi_{lk}=i\varepsilon^{ijk}-\frac{\theta}{2}\left(p_i\delta^{jk}-p_j\delta^{ik}\right)-\frac{i\theta \chi}{12}\varepsilon^{ijk}p^2+\frac{i\theta \chi}{12}\varepsilon^{ijl}p_lp_k.
\end{equation}
Substituting (\ref{27}), (\ref{28}) and (\ref{29}) in (\ref{26}) we conclude that the function $\chi(t)$ has to satisfy  the ODE
\begin{equation}\label{30}
  2t\frac{d\chi}{dt}+3(\chi+1)-\frac{t \chi^2}{6}=0,
\end{equation}
with initial condition $\chi(0)=-1$. The solution is given by
\be\label{30a}
\chi(t)=-\frac{6}{t}\left(\sqrt{\frac{t}{2}}\coth\sqrt{\frac{t}{2}}-1\right).
\ee
In the form of series it can be represented as
\be\label{30b}
\chi(t)=-6\sum_{n=1}^\infty\frac{2^nB_{2n}t^{n-1}}{(2n)!},
\ee
where $B_n$ are Bernoulli numbers, $B_1=-1/2$, $B_2=1/6$, $B_3=0$, etc. In particular, one may see that
\be
\chi(0)=-6\frac{2B_{2}}{2}=-1.
\ee

The closed form of the polydiffferential representation for the algebra (\ref{1}) reads then
\begin{equation}\label{31}
  \hat x^i=x^i+\frac{i\theta}{2}\varepsilon^{ijk}x^k\partial_j+
  (x^i\Delta-x^l\partial_l\partial_i)\Delta^{-1}\left[\frac{\theta}{2}\sqrt{\Delta}\coth\left(\frac{\theta}{2}\sqrt{\Delta}\right)-1\right].
  \end{equation}
Taking into account Eq.  \eqn{30b} we may write Eq. \eqn{31} in the form of an infinite power series,
\begin{equation}\label{31a}
    \hat x^i=x^i+\sum_{n=1}^\infty \left({i\theta}\right)^n \frac{(-1)^nB_n}{n!}x^{k_n}\varepsilon^{ij_1k_1}\varepsilon^{k_1j_2k_2}\dots\varepsilon^{k_{n-1}j_nk_n}\partial_{j_1}\dots\partial_{j_n},
\end{equation}
which is in agreement with  previous results \cite{Dito,Gutt,Meljanac}, obtained within  different approaches.
Let us prove the equivalence between the two expressions.
First we note that \eqn{31a} can be written as
\be\label{g1}
\hat x^i=x^i+\frac{i\theta}{2}\varepsilon^{ijk}x^k\partial_j+x^l\mathcal{X}^i_l\left(-{\theta^2 M}/{2}\right),
\ee
where the operator valued matrix $M^i_l$ is defined as
\be\label{g2}
M^i_l=\varepsilon^{ij_1k_1}\varepsilon^{k_1j_2l}\partial_{j_1}\partial_{j_2}=\partial_i\partial_l-\delta^{il}\Delta,
\ee
the function $\mathcal{X}(t)$ is given by
\be\label{g3}
\mathcal{X}(t)=\sqrt{\frac{t}{2}}\coth\sqrt{\frac{t}{2}}-1=\sum_{n=1}^\infty\frac{2^nB_{2n}t^{n}}{(2n)!},
\ee
and the expression $\mathcal{X}^i_l\left(-{\theta^2 M}/{2}\right)$ is understood as a function of the matrix $M$ in the sense of power series,
 \be\label{g4}
\mathcal{X}^i_l\left(-{\theta^2 M}/{2}\right)=\sum_{n=1}^\infty\frac{(-1)^n\theta^{2n}B_{2n}}{(2n)!}M^i_{l_1}M^{l_1}_{l_2}\dots M^{l_n}_{l}.
\ee
Indeed, taking into account the form of the matrix $M^i_l$, given by \eqn{g2} and the fact that $B_{2n+1}=0$, $n>1$, the expression \eqn{g4} becomes
 \be\label{g5}
 \mathcal{X}^i_l\left(-{\theta^2 M}/{2}\right)=\sum_{n=2}^\infty \frac{\left({i\theta}\right)^nB_n}{n!}\varepsilon^{ij_1k_1}\varepsilon^{k_1j_2k_2}\dots\varepsilon^{k_{n-1}j_nl}\partial_{j_1}\dots\partial_{j_n}
 \ee
 which  is nothing but \eqn{31a}, as stated.

Next we note that the matrix \eqn{g2} is diagonalizable, i.e.,
\be\label{g6}
M=S\cdot D\cdot S^{-1},
\ee
where $D$ is the  diagonal matrix with the eigenvalues of $M$ on the diagonal and the matrix $S$ consists of the corresponding eigenvectors,
\begin{equation*}
     D=\left(
                \begin{array}{ccc}
                  0& 0& 0 \\
                  0 &- \Delta &0 \\
                  0& 0&- \Delta \\
                \end{array}
              \right),\,\,\,\, S=\left(
                             \begin{array}{ccc}
                               \partial_x/\partial_z & -\partial_z/\partial_x & -\partial_y/\partial_x \\
                               \partial_y/\partial_z & 0 & 1 \\
                               1 & 1 & 0 \\
                             \end{array}
                           \right).
     \end{equation*}
With the help of the expression \eqn{g6} we may write \eqn{g4} as
\begin{equation}\label{g7}
\mathcal{X}\left(-{\theta^2 M}/{2}\right)=S\cdot \left(
                \begin{array}{ccc}
                  0& 0& 0 \\
                  0 &\mathcal{X}\left(\frac{\theta^2}{2}\Delta\right) &0 \\
                  0& 0&\mathcal{X}\left(\frac{\theta^2}{2}\Delta\right) \\
                \end{array}
              \right)\cdot S^{-1}.
\end{equation}
Finally we obtain,
\be\label{g8}
x^l\mathcal{X}^i_l\left(-{\theta^2 M}/{2}\right)=(x^i\Delta-x^l\partial_l\partial_i)\Delta^{-1}\left[\frac{\theta}{2}\sqrt{\Delta}\coth\left(\frac{\theta}{2}\sqrt{\Delta}\right)-1\right],
\ee
which after substitution in \eqn{g1} reproduces exactly the closed formula \eqn{31}. That is, using the property of the matrix \eqn{g2} to be diagonalizable and the form of the function \eqn{g3} one may represent the formal series \eqn{31a} as a closed expression.

Let us  stress that the formula \eqn{31} is an important step in our construction, since it will be used to obtain the closed expression for the gauge operator $T$ mentioned in the introduction. Also we note that  even if Eq. \eqn{31} concerns $\mathfrak{su}(2)$ generators, closed expressions for a wide class of other Lie algebra generators can be  easily obtained following the same approach.

\section{Generalization}
The above construction can be generalized to any Lie algebra $\mathfrak{g}$, with commutation relations,
 \begin{equation}\label{Lie}
    \left[ \hat{x}^{i},\hat{x}^{j}\right] =i \theta {f} ^{ij}_k\hat{x}^{k},\,\,\,i,j,k=1,\dots d,
 \end{equation}
provided that the corresponding matrix
 \be\label{g7}
 M_\mathfrak{g}^{il}={f}^{ij_1}_k {f}^{kj_2}_l\partial_{j_1}\partial_{j_2},
 \ee
 be diagonalizable. That is, if there exists a nondegenerate matrix $S_\mathfrak{g}$, such that $M_\mathfrak{g}=S_\mathfrak{g}\cdot D_\mathfrak{g}\cdot S_\mathfrak{g}^{-1}$, where $ D_\mathfrak{g}$ is diagonal. As in the case of the $\mathfrak{su}(2)$ algebra we  first write
 \be\label{g8}
 \hat x^i=x^l g^i_l(\partial)=x^l\left[\delta^{il}+\frac{i\theta}{2}{f}^{ij}_l\partial_j+\mathcal{X}^i_l\left(-{\theta^2}M_\mathfrak{g}/{2}\right)\right].
 \ee
 Taking into account \eqn{g4} and \eqn{g7} the above expression can be written as,
 \begin{equation}\label{g9}
    \hat x^i=x^i+\sum_{n=1}^\infty \left({i\theta}\right)^n \frac{(-1)^nB_n}{n!}x^{k_n}{f}^{ij_1}_{k_1}{f}^{k_1j_2}_{k_2}\dots{f}^{k_{n-1}j_n}_{k_n}\partial_{j_1}\dots\partial_{j_n},
\end{equation}
which corresponds to the formal series giving the polydifferential representation for the generators of the algebra \eqn{Lie}, see \cite{Gutt,Dito,Meljanac}. On the other hand, the fact that $M_\mathfrak{g}$ is diagonalizable implies that
\be\label{g10}
\mathcal{X}^i_l\left(-{\theta^2}M_\mathfrak{g}/{2}\right)=\left[S_\mathfrak{g}\cdot \mathcal{X}\left(-{\theta^2}D_\mathfrak{g}/{2}\right)\cdot S_\mathfrak{g}^{-1}\right]^{i}_l,
\ee
where
 \be
 \mathcal{X}\left(-{\theta^2}D_\mathfrak{g}/{2}\right)=\rm{diag}\left[ \mathcal{X}\left(-{\theta^2}\lambda_1/{2}\right),\dots,\mathcal{X}\left(-{\theta^2}\lambda_d/{2}\right)\right],
 \ee
 with $\lambda_1,\dots,\lambda_d$ being the differential operators corresponding to eigenvalues of \eqn{g7}. That is, \eqn{g8} becomes
 \be\label{g11}
 \hat x^i=x^i+\frac{i\theta}{2}{f}^{ij}_l x^l\partial_j+x^l\left[S_\mathfrak{g}\cdot \mathcal{X}\left(-{\theta^2}D_\mathfrak{g}/{2}\right)\cdot S_\mathfrak{g}^{-1}\right]^{i}_l.
 \ee
 This equation represents  the closed expression for the polydifferential representation of \eqn{Lie} which, as we will see, is compatible with the construction of the symmetrically ordered star product on the dual of the appropriate Lie algebra $\mathfrak{g}$.

\section{Weyl star product for linear Poisson structures}\label{weyl}

On using the polydifferential representation \eqn{g11} we shall construct in this section  the symmetrically ordered star product corresponding to a generic Lie algebra \eqn{Lie}. First we observe that since the terms proportional to n-th powers of  $\theta, n\ge 1$  in (\ref{g11}), or what is the same, \eqn{g9}, contain derivatives and  $\partial_i\triangleright  1=0$, we conclude that
\begin{equation*}
   \hat x^i\triangleright 1=x^i.
\end{equation*}
Also one may easily verify that the action of the symmetrized product of two coordinate operators on constants gives as a result the product of the corresponding coordinates,
\begin{equation*}
    \langle\hat x^i\hat x^j\rangle_W\triangleright 1=\frac{1}{2}\left(\hat x^i\hat x^j+\hat x^j\hat x^i\right)\triangleright 1=x^ix^j
\end{equation*}
where the symbol $\langle\dots\rangle_W$ stands for the symmetric or Weyl ordering of operators. The construction \cite{KV} of the  polydifferential representation (\ref{g8}) for the algebra (\ref{Lie}) guarantees that the above property holds true for the symmetrized product of any number of coordinate operators, i.e.,
\begin{equation}\label{32}
 \langle\hat x^{i_1}\dots\hat x^{i_n}\rangle_W\triangleright 1= \frac
1{n!}\sum_{P_n} P_n(\hat x^{i_1}\dots \hat x^{i_n})\triangleright 1=x^{i_1}\dots x^{i_n}.
\end{equation}
where $P_n$ are all permutations of $n$ elements. Eq. (\ref{32}) is the key difference between the representation (\ref{g11})   and the ones including only a finite number of derivative terms  see e.g., \cite{GLMV,RV}.

Suppose that a function $f(x)$ on $\mathbb{R}^3$ can be expanded in
the Taylor series around zero,
\begin{equation}
f(x)=\sum_{n=0}^\infty f^{(n)}_{i_1\dots i_n}x^{i_1}\dots x^{i_n}.
\label{33}
\end{equation}
We define the Weyl ordered operator $f_W(\hat x)$ corresponding to the function $f(x)$ by the following rule,
\begin{equation}
f_W(\hat x)=\sum_{n=0}^\infty f^{(n)}_{i_1\dots i_n} \sum_{P_n} \frac
1{n!} P_n(\hat x^{i_1}\dots \hat x^{i_n}).  \label{34}
\end{equation}
Since $\exp[i k_m \hat x^m]$ is automatically Weyl ordered,  we may also use the Weyl map as another way of implementing  the symmetric ordering
\begin{equation}
f_W(\hat x)\equiv \hat W(f)  =\int \frac{d^{3}k}{\left( 2\pi
\right) ^{3}}\tilde{f}\left( k\right) e^{ik_{m}\hat{x}^{m}},  \label{35}
\end{equation}
where $\tilde{f}\left( k\right)$ is the Fourier transform of $f$. Equations (\ref{32}) and (\ref{34}) imply that
\begin{equation}\label{36}
    \hat W(f)\triangleright 1=f(x).
\end{equation}
This in turn means that the inverse Weyl map, $W^{-1}$, in this case is nothing but the action of the corresponding operator on the constant,
\begin{equation}\label{37}
   W^{-1}\left(f_W(\hat x)\right)=f_W(\hat x)\triangleright 1=f(x),
\end{equation}
see \cite{KV} and \cite{Meljanac2} for more details.

Let us define the star product as
\begin{equation}
\hat W(f\star_W g)=\hat W(f) \hat W(g).  \label{38}
\end{equation}
By construction this star product is associative due to the associativity of the operator product.
Taking into account Eq. (\ref{37}) we write
 \begin{eqnarray}
&&(f\star_W g)(x)=W^{-1}\left(\hat W(f) \hat W(g)\right)\label{39}\\
&&=\hat W\left( f\right) \hat W\left( g\right)\triangleright 1= \hat W\left( f\right)\triangleright g(x)
= \int \frac{d^{3}k}{\left( 2\pi
\right) ^{3}}\tilde{f}\left( k\right) e^{ik_{m}\hat{x}^{m}} \triangleright g(x)~,\notag
\end{eqnarray}%
where the RHS  represents the action of a polydifferential operator
on a function.

The property (\ref{32}) ensures that Eq. (\ref{39}) satisfies  the condition
\begin{equation}\label{40}
    \frac
1{n!}\sum_{P_n} P_n( x^{i_1}\star_W \dots \star_W x^{i_n})=x^{i_1}\dots x^{i_n}.
\end{equation}
Also, by  construction the star product $\star_W$ is Hermitean, i.e.,
\begin{equation}\label{41}
    (f\star_W g)^\ast=g^\ast \star_W f^\ast.
\end{equation}

The star product between the coordinate $x^i$ and a function $f(x)$ is given by $x^i\star_W f=\hat x^i\triangleright f$, where the closed formula for the operator $\hat x^i$ was calculated in \eqn{g11}, or \eqn{31} in case of $\mathfrak{su}(2)$.
For the latter case we have in particular
\beqa
x^i\star_W f &=&\left\{x^i+\frac{i\theta}{2}\varepsilon^{ijk}x^k\partial_j\right.\\
&+&\left.
  (x^i\Delta-x^l\partial_l\partial_i)\Delta^{-1}\left[\frac{\theta}{2}\sqrt{\Delta}\coth\left(\frac{\theta}{2}\sqrt{\Delta}\right)-1\right]\right\}\triangleright f.\notag \label{poly}
\eeqa
The star product
 \begin{equation}\label{Xstar}
   ( x^{i_1}\dots x^{i_k})\star_W f(x)=\langle\hat x^{i_1}\dots\hat x^{i_k}\rangle_W\triangleright f(x),
\end{equation}
can be constructed by induction using the formula \cite{Gutt},
\begin{equation}\label{comb}
    \langle\hat x^{i_1}\dots\hat x^{i_k}\rangle_W=\frac{1}{k}\sum_{l=1}^k \hat x^{i_l}\langle\hat x^{i_1}\dots\check x^{i_l}\dots \hat x^{i_k}\rangle_W,
\end{equation}
where $\check x^{i_l}$ denotes omission.
In particular
\begin{eqnarray}
 && (x^{i_1}x^{i_2})\star_W f= \frac{1}{2}\left(\hat x^{i_1}\hat x^{i_2}+\hat x^{i_2}\hat x^{i_1}\right)\triangleright f\\
&& =\left[ x^{l_1}x^{l_2}g_{i_1l_1}g_{i_2l_2}+\frac{1}{2}x^{l_1}(g_{i_2l_2}\partial^{l_2}g_{i_1l_1}+ g_{i_1l_2}\partial^{l_2}g_{i_2l_1})\right]\triangleright f \notag
\end{eqnarray}

We will need the expression for the star product in the form of the infinite series,
\begin{equation}\label{Wstar}
    (f\star_W g)(x)=f\cdot g+\sum_{n=1}^\infty \left(\frac{i\theta}{2}\right)^nC_n(f,g),
\end{equation}
where $C_n(f,g)$ are bidifferential operators. {Usually the operator $C_1(f,g)$ is fixed from the requirement $C_1(g,h)$-$C_1(h,g)=\{f,g\}$, where $\{f,g\}$ is the Kirillov-Poisson bracket corresponding to the algebra \eqn{Lie}, while the higher order operators $C_n(f,g)$ are obtained from the requirement of the associativity of the star product. In our case the associativity of \eqn{Wstar} follows from the definitions \eqn{38}, \eqn{39}, which in turn are a consequence of the Jacobi identity for the Lie algebra structure constants. Also from the construction it follows that} each term $C_n(f,g)$ should contain the product of $n$ linear Poisson bi-vectors $\omega^{ij}=f^{ij}_kx^k$ contracted with $2n$ derivatives acting on functions $f$ and $g$ and $(\omega)^n$. The property \eqn{41} implies that
\begin{equation}\label{43}
    C_n(f,g)=(-1)^n C_n(g,f).
\end{equation}
 The general form of bidifferential operator $C_n(f,g)$, satisfying the above requirements is
\begin{eqnarray}\label{42}
&& C_n(f,g)=\sum_{l=1}^n \sum_{a=l}^n C^n_{a,a-l+1} x^{k_1}\dots x^{k_l}\underbrace{{f}^{i_1j_1}_{k_1}\dots{f}^{i_lj_l}_{k_l}}_{l}\cdot\\
&&
\underbrace{{f}^{i_{l+1}j_{l+1}}_{i_n}{f}^{i_{l+2}j_{l+2}}_{i_{n-1}}\dots{f}^{i_{l+n-a}j_{l+n-a}}_{i_{a+1}}}_{n-a}
\underbrace{{f}^{i_{l+n-a+1}j_{l+n-a+1}}_{j_{a-l}}\dots{f}^{i_nj_n}_{j_1}}_{a-l}\nonumber\\
&&\left(\partial_{i_1}\dots\partial_{i_a}f\partial_{j_{a-l+1}}\dots\partial_{j_n}g+(-1)^n\partial_{i_1}\dots
\partial_{i_a}g\partial_{j_{a-l+1}}\dots\partial_{j_n}f\right),\nonumber
\end{eqnarray}
where coefficients $C^n_{m,l}$, with $l\leq m\leq n$ can be determined using \eqn{Xstar}. In particular, from the expression $x^i\star_W f=\hat x^i\triangleright f$ and formula \eqn{g9} we conclude that
\begin{equation}\label{44}
    C^n_{n,n}=\frac{1}{n!}B_n.
\end{equation}
The coefficient  $C^n_{n,1}=1/n!$, etc.
By  construction, \eqn{42} obey the condition \eqn{43}.

Up to the third order one writes then
\begin{eqnarray}
&&f\star_W g =f\cdot g+\frac{i\theta}{2}x^{k_1}{f}^{{i_1}{j_1}}_{{k_1}}\partial_{i_1} f\partial_{j_1} g\label{16} \\&&-\frac{\theta ^{2}}{4}\left[\frac{1}{4}x^{k_1}x^{k_2}{f} ^{{i_1}{j_1}}_{{k_1}}{f} ^{{i_2}{j_2}}_{{k_2}}(\partial _{i_1}\partial
_{i_2}f\partial _{j_1}\partial _{j_2}g+\partial _{i_1}\partial
_{i_2}g\partial _{j_1}\partial _{j_2}f)\right.  \notag \\
&&-\left.\frac{1}{3}x^{k_1}{f}^{{i_1}{j_1}}_{{k_1}}{f} ^{{i_2}{j_2}}_{{j_1}}\left( \partial _{i_1}\partial _{i_2}f\partial _{j_2}g+\partial
_{i_1}\partial _{i_1}g\partial _{j_2}f\right)\right]\notag \\&&-\frac{i\theta^3}{8}\left[\frac{1}{6}x^a{f}^{nk}_{a}{f} ^{jm}_{n}{f}
^{il}_{m}\left( \partial _{i}\partial _{j}f\partial _{k}\partial _{l}g-\partial
_{i}\partial _{j}g\partial _{k}\partial _{l}f\right) \right.\notag\\
&&+\frac{1}{3}x^ax^b{f} ^{ln}_{a}{f} ^{ik}_{b}{f} ^{jm}_{l}\left(
\partial _{i}\partial _{j}f\partial _{k}\partial _{n}\partial _{m}g-\partial
_{i}\partial _{j}g\partial _{k}\partial _{n}\partial _{m}f\right)
\notag \\
&&\left. +\frac{1}{6}x^ax^bx^c{f} ^{jl}_{a}{f} ^{im}_{b}{f} ^{kn}_{c}\partial _{i}\partial
_{j}\partial _{k}f\partial _{l}\partial _{n}\partial _{m}g  \right]+O\left(\theta^4\right)
~.  \notag
\end{eqnarray}%

To obtain the  Weyl star product  in closed form one may use the Baker-Campbel-Hausdorff (BCH) formula. The combination of the definition of the star product \eqn{38} with the Weyl map \eqn{35} implies
\be\label{w1}
f\star_W g(x) =\int \frac{d^{d}k_1}{\left( 2\pi
\right) ^{d}} \frac{d^{d}k_2}{\left( 2\pi
\right) ^{d}}\tilde{f}\left( k_1\right)\tilde{g}\left( k_2\right)W^{-1}\left( e^{ik_1^{m}\hat{x}^{m}}e^{ik_2^{m}\hat{x}^{m}}\right).
\ee
Then the expression $W^{-1}\left( e^{ik_1^{m}\hat{x}^{m}}e^{ik_2^{m}\hat{x}^{m}}\right)$ can be calculated using the BCH formula, see e.g. \cite{Behr}. If the coordinate operators $\hat x^i$ satisfy the Lie algebra commutation relations, then
\be\label{w2}
W^{-1}\left( e^{ik_1^{m}\hat{x}^{m}}e^{ik_2^{m}\hat{x}^{m}}\right)=e^{iB_m(k_1,k_2){x}^{m}}.
 \ee
where the function $B_m(k_1,k_2)$ depends on the structure constants of the corresponding Lie algebra and satisfies the following properties,
 \be\label{B0}
B_m(k,0)=B_m(0,k)=k_m,
\ee
which is a consequence of the stability of the unity, i.e., $f\star_W 1=1\star_Wf=f(x)$, and the condition
\be
B_m(k_1,k_2)=-B_m(-k_2,-k_1),
\ee
coming from the equation \eqn{41}, i.e., hermiticity of the Weyl star product.

In the form of an infinite series the function $B_m(k_1,k_2)$ is given by the Dynkin formula. For some specific Lie algebras its closed form  is known.
In case of $\mathfrak{su}(2)$ one has, see \cite{FM,OR13},
 \be\label{Bk}
\vec B(k_1,k_2)=\left.\frac{2\arcsin\left|\vec p_1\oplus\vec p_2\right|}{\theta\left|\vec p_1\oplus\vec p_2\right|}\vec p_1\oplus\vec p_2\right|_{ \vec p_a=\vec k_a\sin\left(\frac{\theta}{2}|k_a|\right)/|k_a|},
\ee
where $a=1,2$ and
\be\label{p1p2}
\vec p_1\oplus\vec p_2=\sqrt{1-| p_1|^2} \vec p_2+\sqrt{1-| p_2|^2}\vec p_1-\vec p_1\times\vec p_2.
\ee

To verify  that the formulas \eqn{39} and \eqn{w1}, with $B_m(k_1,k_2)$ defined in \eqn{Bk}, correspond to the same star product we will calculate $x^i\star_Wf$ using \eqn{w1} and then compare with the expression $\hat x^i\triangleright f$, where the operator $\hat x^i$ is given by \eqn{31}. Since the Fourier transform of $x^i$ is the derivative of a Dirac delta function, we get
\beqa
x^i\star_Wf &=&\int \frac{d^{3}k_1}{\left( 2\pi
\right) ^{3}} \frac{d^{3}k_2}{\left( 2\pi
\right) ^{3}}(2\pi i )^3\left(\partial^i_{k_1}\delta(k_1)\right)\tilde{f}\left( k_2\right)e^{iB_m(k_1,k_2)x^m}\\
&=&-\int \frac{d^{3}k_1}{\left( 2\pi
\right) ^{3}} d^{3}k_2 x^l\frac{ \partial B_l(k_1,k_2)}{\partial k_1^i}\delta(k_1)\tilde{f}\left( k_2\right)e^{iB_m(k_1,k_2)x^m}\notag\\
&=&-\int \frac{d^{3}k_2}{\left( 2\pi
\right) ^{3}}  x^l\left.\frac{ \partial B_l(k_1,k_2)}{\partial k_1^i}\right|_{k_1=0}\tilde{f}\left( k_2\right)e^{iB_m(0,k_2)x^m}\notag
\eeqa
After some algebra one may see that
\beqa
&&\left.\frac{ \partial B_l(k_1,k_2)}{\partial k_1^i}\right|_{k_1=0}=\\
&&-\theta \varepsilon^{ilm}k_2^m+\delta^{il}\frac{\theta}{2}|k_2|\cot\left(\frac{\theta}{2}|k_2|\right)
+\frac{k_2^ik_2^l}{|k_2|^2}\left(\frac{\theta}{2}|k_2|\cot\left(\frac{\theta}{2}|k_2|\right)-1\right).\notag
\eeqa
Then, taking into account \eqn{B0} and integrating over $k_2$ we conclude that indeed $x^i\star_Wf=\hat x^i\triangleright f$.

\section{Trace functional and equivalent star products}\label{gaugeop}

In this section we will describe the perturbative scheme for the construction of the gauge operator $T$ which transforms a given star product $\star'$, corresponding to the Poisson bi-vector $\omega^{ij}(x)$ to the one satisfying the closed property (\ref{2}). For the moment we do not specify the star product,  since, as we will see, the procedure is applicable for any one.

We will start introducing  the following notations,
\begin{equation}\label{star}
    f\star' g=f\cdot g+\sum_{n=1}^\infty \theta^n C'_n(f, g),
\end{equation}
where $\theta$ is the deformation parameter,
\begin{equation}
    C'_1(f,g) =\frac{i}{2}\{f,g\}=\frac{i}{2}\partial_i f\omega^{ij}\partial_j g
\end{equation}
and  $C'_n(f, g)$ stands for  bi-differential operators of higher degree.
Further, following \cite{Felder,Pinzul}, let us define for any given function $f$
\begin{equation}\label{trace}
     \Tr\left( f\right)=\int\, \dd^Nx\,\mu(x)\cdot f(x)
\end{equation}
with $\mu(x)$ being some  weight function satisfying
\begin{equation}\label{weight}
   \partial_i(\mu \omega^{ij})=0.
\end{equation}
Let's now check  the closed property for the  star product (\ref{star}), integrating by parts the LHS of Eq.  (\ref{2}). At  first order in $\theta$ this condition holds true because of the property  (\ref{weight}) of the weight function $\mu(x)$. But already at  second order this condition might be violated, as  it happens for example with the star product in \cite{KV}, see \cite{Kup15} for details.

To solve this problem we search for a gauge transformation, $T$, in the form of a perturbative expansion in $\theta$
 \begin{equation}
    T=1+\theta T_1+\theta^2T_2+ ...\,,
\end{equation}
which maps the star product $\star'$ to the new one, $\star$, according to Eq. \eqn{gauge}, satisfying the desired property (\ref{2}). From Eq.  (\ref{gauge}) we have
\begin{equation}\label{D}
  f \star g=T^{-1}\left(Tf\star' T g\right)
\end{equation}
and we require that
\begin{equation}\label{3}
    \Tr\left(  f \star g\right)- \Tr\left(  f \cdot g\right)=0.
\end{equation}
Thus, the gauge operator $T$ has to satisfy the following
\begin{equation}\label{C}
   \Tr\left[  T^{-1}\left(T f\star' T g\right)-f \cdot g\right]=0.
\end{equation}
Once the operator $T$ has been found, the new star product $\star$ can be recovered by the formula (\ref{D}).

Let us discuss the condition (\ref{C}) in details. Our aim here is just to figure out the procedure of the construction of the operator $T$. We will not provide the explicit formula for it. First we note that since,
\begin{equation*}
     Tr\left[  C'_1(f, g)\right]=0,
\end{equation*}
it is reasonable to set, $T_1=0$, i.e.,
\begin{equation}\label{4}
    T=1+\sum_{n=2}^\infty \theta^n T_n.
\end{equation}
For the inverse operator we write,
\begin{equation}\label{5}
    T^{-1}=1+\sum_{n=2}^\infty \theta^n \bar T_n.
\end{equation}
From the equation
\begin{equation*}
T^{-1}\circ T=1,
\end{equation*}
 one finds that $\bar T_2=-T_2$, $\bar T_3=-T_3$, $\bar T_4=T^2_2-T_4$, etc. Therefore,
 \begin{equation}\label{6}
    T^{-1}=1-\theta^2T_2-\theta^3T^3+\theta^4\left(T^2_2-T_4\right)+\dots\,.
 \end{equation}
Let us write,
\begin{eqnarray}
&&T^{-1}\left(T f \star' Tg\right)=f\cdot g +\theta C'_1(f, g)\label{7} \\
&&+\theta^2\left[T_2f\cdot g+f\cdot T_2 g -T_2(f\cdot g)+C'_2(f, g)\right]\nonumber\\
&&+\theta^3\left[T_3f\cdot g+f\cdot T_3 g -T_3(f\cdot g)\right.\nonumber\\
&&+\left.C'_1(T_2f, g)+C'_1(f,T_2 g)  -T_2C'_1(f, g)+C'_3(f, g)\right]+\dots\,.\nonumber
\end{eqnarray}
Substituting the above equation in the condition (\ref{C}) we obtain at each order in $\theta$ the following equations for the operators $T_n$:
\begin{eqnarray}
&&\Tr\left[T_2(f\cdot g)-T_2f\cdot g-f\cdot T_2 g\right]= \Tr\left[C'_2(f, g)\right]\,, \label{8} \\
&&\Tr\left[T_3(f\cdot g)-T_3f\cdot g-f\cdot T_3 g\right] \nonumber\\
&&=\Tr\left[C'_3(f, g)+C'_1(T_2f, g)+C'_1(f, T_2 g)  -T_2C'_1(f, g)\right]\,,\nonumber\\
&&\ldots\nonumber\\
&&\Tr\left[T_n(f\cdot g)-T_nf\cdot g-f\cdot T_n g\right]=\Tr\left[C'_n(f, g)+ \rm{lower\, order\, terms}\right]\,.\nonumber
\end{eqnarray}
Since the star product $\star'$ at the  $n$-th order in  $\theta$ contains derivatives of degree not higher than $n$, and  one may integrate by parts the RHS  of Eqs. (\ref{8}) as many times as necessary  to lower the order of these derivatives,
we make the following ansatz for the operators $T_n$:
\begin{equation}\label{Dn}
    T_n=b^0_n(x)+b^{ i_1}_n(x)\partial_{i_1}+...+b^{ i_1i_2...i_n}_n(x)\partial_{i_1}\partial_{i_2}...\partial_{i_n}
\end{equation}
that is, it should be a differential operator of  order not higher than $n$ with  coefficients in the algebra of functions. Note that, by  construction, the coefficient functions $b^{ i_1i_2...i_k}_n(x)$ must be symmetric in all indices.  Eqs.  (\ref{8}) will result in a set of  algebraic equations for the coefficient functions $b^{ i_1i_2...i_k}_n(x)$. Solving these equations one finds the operator $T$.

To illustrate our procedure we will construct the gauge operator in the first nontrivial order, $\theta^2$, for the star product
\begin{align}
& (f\star_W g)(x)=f\cdot g+\frac{i\theta }{2}\partial
_{i}f\omega ^{ij}\partial _{j}g -\frac{\theta ^{2}}{8}\omega ^{ij}\omega ^{kl}\partial
_{i}\partial _{k}f\partial _{j}\partial _{l}g \label{star1} \\
& -\frac{\theta^2}{12}\omega
^{ij}\partial _{j}\omega ^{kl}\left( \partial _{i}\partial _{k}f\partial
_{l}g-\partial _{k}f\partial _{j}\partial _{l}g\right)  +O\left(
\theta ^{3}\right)   \notag
\end{align}
with $\star_W= \star'$.
Integrating by parts the term $\Tr\left[C'_2(f, g)\right]$ in the first of Eqs. \eqn{8} we end up with
\begin{equation}
\Tr\left[C'_2(f, g)\right]=\frac{1}{24}\int \dd^{N}x\partial_{i}f\partial_{l}\left(  \mu
\omega^{ij}\partial_{j}\omega^{lk}\right)  \partial_{k}g; \label{9}%
\end{equation}
that is, the operator $T_2$ can be chosen in this case to be
\begin{equation}\label{10}
    T_2=b_2^{ik}(x)\partial_i\partial_k\,.
\end{equation}
Then we compute
\begin{equation}\label{11}
    T_2(f\cdot g)-T_2f\cdot g-f\cdot T_2 g=2b_2^{ik}(x)\partial_if\partial_kg\,.
\end{equation}
Substituting Eqs. (\ref{9}) and (\ref{11}) in Eq. (\ref{8}) one finds the algebraic equation
\begin{equation}\label{12}
    2\mu b_2^{ik}=\frac{1}{24}\partial_{l}\left(  \mu
\omega^{ij}\partial_{j}\omega^{lk}\right).
\end{equation}
Note that since the coefficient $b_2^{ik}$ should be symmetric in the indices $ik$, one finds the following consistency condition for the existence of the gauge operator in this case:
\begin{equation}\label{13}
    \partial_{l}\left(  \mu\omega^{ij}\partial_{j}\omega^{lk}\right)
-\partial_{l}\left(  \mu\omega^{kj}\partial_{j}\omega^{li}\right)=0.
\end{equation}
However, due to the Jacobi identity on the Poisson structure $\omega^{ij}(x)$ and the definition (\ref{weight}) of the weight function $\mu$, one may see that
\begin{eqnarray*}
     &&\partial_{l}\left(  \mu\omega^{ij}\partial_{j}\omega^{lk}
+  \mu\omega^{kj}\partial_{j}\omega^{il}\right)= - \partial_{l}\left(  \mu\omega^{lj}\partial_{j}\omega^{ki}\right)\\
&&
=- \partial_{l}\left(  \mu\omega^{lj}\right)\partial_{j}\omega^{ki}-  \mu\omega^{lj}\partial_{l}\partial_{j}\omega^{ki}=0.
\end{eqnarray*}
That is, (\ref{13}) holds true and the solution of (\ref{12}) is given by
\begin{equation}\label{14}
    b_2^{ik}=\frac{1}{48\mu}\partial_{l}\left(  \mu
\omega^{ij}\partial_{j}\omega^{lk}\right).
\end{equation}
The modified star product takes the form,
\begin{align}
& (f\star g)(x)=f\cdot g+\frac{i\theta }{2}\partial
_{i}f\omega ^{ij}\partial _{j}g -\frac{\theta ^{2}}{8}\omega ^{ij}\omega ^{kl}\partial
_{i}\partial _{k}f\partial _{j}\partial _{l}g \label{star2} \\
& -\frac{\theta^2}{12}\omega
^{ij}\partial _{j}\omega ^{kl}\left( \partial _{i}\partial _{k}f\partial
_{l}g-\partial _{k}f\partial _{j}\partial _{l}g\right) -\frac{\theta ^{2}}{24\mu}\partial_{l}\left(  \mu
\omega^{ij}\partial_{j}\omega^{lk}\right) \partial_{i}f \partial_{k}g +O\left(
\theta ^{3}\right) .  \notag
\end{align}

Note that if the weight function $\mu(x)$ is a constant, the last term in the second line of the above equation becomes just
\begin{equation*}
\frac{\theta ^{2}}{24}\partial_{l}
\omega^{ji}\partial_{j}\omega^{lk} \partial_{i}f \partial_{k}g.
\end{equation*}
That is, in this case at least up to the second order in $\theta$ the closed star product \eqn{star2} is nothing but the Kontsevich star product, see e.g., in \cite{DS}. This observation is in agreement with general result proved in \cite{Felder}, stating that for a Poisson bivector field $\omega^{ij}\left( x\right)$ and a constant volume forme ${\bf\Omega}$, such that $\mathrm{div}_\Omega \omega=0$, the Kontsevich star product constructed from $\omega$ is closed.

For some Lie algebras, like e.g., $\mathfrak{su}(2)$,  the weight function $\mu(x)$ can be set to be a constant. Therefore the corresponding closed star product is necessarily the Kontsevich one. For others, like $[\hat x,\hat y]=i\theta \hat x$, the function $\mu(x)$ is not constant and the expression for both, gauge operator $T$, and the closed star product will depend on $\mu(x)$ and its derivatives. The gauge operator relating the Kontsevich and the Gutt star products on the dual of a Lie algebra was constructed in \cite{Dito} as a formal power series involving the Kontsevich weights. In the next section we will derive the closed formula for the gauge operator $T$ for the algebra $\mathfrak{su}(2)$.

\section{Closed star product on $\mathbb{R}^3_\theta$}\label{gaugeweyl}

Now let us come to  the specific  case of the noncommutative algebra $\mathbb{R}^3_\theta$.
Since the coordinate operators satisfy the algebra (\ref{1}), the corresponding Poisson structure is just $\omega^{ij}=\varepsilon^{ijk}x^{k}$.
One may see that any function on $r^{2}=x^{2}+y^{2}+z^{2}$ can be
chosen as a weight function $\mu\left(  x\right)  $ in the definition of trace (\ref{trace}), since
\begin{equation*}
    \partial_{i}\left( \mu\left(  r^{2}\right) \varepsilon^{ijk}x^{k}  \right)  =0.
\end{equation*}
 For simplicity we set
$\mu  =1$, i.e., the trace functional now is just the integration over the space, $ \Tr\left( f\right)=\int\, \dd^3x\, f(x).$

In the previous section we have already calculated the gauge operator and the modified star product up to the second order in $\theta$ for the Weyl star product \cite{KV} corresponding to an arbitrary Poisson structure $\omega^{ij}(x)$. Here we will use this result setting $\omega^{ij}=\varepsilon^{ijk}x^{k}$.  From Eq. \eqn{14} we have then:
\begin{equation}\label{D2}
    T_2=\frac{1}{24}\Delta,
\end{equation}
where $\Delta$ is the ordinary Laplacian.

To find the third order term $T_3$ we have to  solve the equation:
\begin{eqnarray}
&&\int\, \dd^3x\,\left[T_3(f\cdot g)-T_3f\cdot g-f\cdot T_3 g\right]\label{15}\\
&&=\int\, \dd^3x\,\left[f \star_3 g+T_2f\star_1 g+f\star_1 T_2 g -T_2(f\star_1 g)\right]\,,\nonumber
\end{eqnarray}
where, according to  Sec. \ref{weyl} the third order contribution to the Weyl star product corresponding to the algebra (\ref{1}) is given by \eqn{16}.
Integrating by parts the RHS of  Eq. (\ref{15}) we obtain, up to constants
\begin{eqnarray}
&&  \int\, \dd^3x\,\varepsilon^{nkl}\varepsilon ^{jmn}\varepsilon
^{ilm}\left( \partial _{i}\partial _{j}f\partial _{k}g-\partial _{i}\partial _{j}g\partial _{k}f\right)\label{17}\\
&&=\int\, \dd^3x\,\varepsilon^{nkl}\left(-\delta^j_i\delta^n_l+\delta^j_l\delta^n_i\right)\left( \partial _{i}\partial _{j}f\partial _{k}g-\partial _{i}\partial _{j}g\partial _{k}f\right)=0.\notag
\end{eqnarray}
That is the RHS  of Eq. (\ref{15}) vanishes after integration by parts. Therefore we can choose
\begin{equation}\label{18}
  T_3=0.
\end{equation}
Thus, up to third order in $\theta$ we obtain
\begin{equation}\label{19}
  T=1+\frac{\theta^2}{24}\Delta+O\left(\theta^4\right).
\end{equation}

Order by order calculation of the operator $T$ and also the form \eqn{40}, \eqn{42} of the Weyl star product indicate that odd powers of $\theta$ are zero and  $T=T(\theta^2\Delta/2)$. This already implies that the operator $\tilde x^i$, defined by
\begin{eqnarray}\label{46}
&&\tilde x^i\triangleright f=    x^i\star f=T^{-1}\left(T(x^i)\star_W Tf\right)\\
&&=T^{-1}\left(x^i\star_W Tf\right)=\left(T^{-1}\circ\hat x^i \circ T\right)f(x),\notag
\end{eqnarray}
can be represented in the form
\begin{equation}\label{47}
    \tilde x^i =\hat x^i+\xi^i(\partial),
\end{equation}
where the operator $\xi^i(\partial)=-2T^{-1} T'\partial_i$, and $T'$ stands for the ordinary derivative with respect to the argument. This is true because
\begin{eqnarray*}
&&T^{-1}\circ\hat x^i \circ T=T^{-1}(\Delta)x^lg_{il}(\partial) T(\Delta)=x^lT^{-1}g_{il} T+\left[T^{-1},x^l\right]g_{il} T\\
&&=\hat x^i-2T^{-1}T'g_{il}\partial_l=\hat x^i-2T^{-1}T'\partial_i.
\end{eqnarray*}
One may easily verify that the operators \eqn{47} satisfy the $\mathfrak{su}(2)$ algebra \eqn{su2}.
The operator $\hat x^i$ is given by \eqn{31}, while $\xi^i(\partial)$ can be found from the condition,
\begin{equation}\label{48}
 \int\, \dd^3x\, \left(x^i\star f-x^i\cdot f\right)=0.
 \end{equation}
 Let us calculate
 \begin{eqnarray}\label{49}
&& \int\, \dd^3x\, \left(\hat x^i\triangleright f-x^i\cdot f\right)=-\theta\int\, \dd^3x\, x^l\phi_{il}(\partial) f\\
&&=-\frac{\theta^2}{12}\int\, \dd^3x\,\partial_l\left[(x^i\partial_l-x^l\partial_i)\chi\left(\frac{\theta^2\Delta}{2}\right)f\right]-\frac{\theta^2}{6}\int\, \dd^3x\,\chi\left(\frac{\theta^2\Delta}{2}\right)\partial_if,\notag
\end{eqnarray}
where the function $\chi(t)$ was defined in sec. 2.
The first term in the second line of \eqn{49} vanishes as a total derivative, while the second one does not. So, if we put
\begin{equation}\label{50}
\xi^i=-2\Delta^{-1}\left[\frac{\theta}{2}\sqrt{\Delta}\coth\left(\frac{\theta}{2}\sqrt{\Delta}\right)-1\right]\partial_i,
\end{equation}
in \eqn{47}, then \eqn{48} holds true. The expression \eqn{46} now becomes
\begin{eqnarray}\label{47a}
 &&  x^i \star f=\left\{x^i+\frac{i\theta}{2}\varepsilon^{ijk}x^k\partial_j\right.\\ &&+\left.(x^i\Delta-x^l\partial_l\partial_i-\partial_i)
   \Delta^{-1}\left[\frac{\theta}{2}\sqrt{\Delta}\coth\left(\frac{\theta}{2}\sqrt{\Delta}\right)-1\right]\right\}\triangleright f(x).\notag
\end{eqnarray}

On the other hand, it implies the differential equation on the function $T(t)$, corresponding to the gauge operator $T$,
\begin{equation}\label{51}
T^{-1} T'=\frac{1}{12}\chi,
\end{equation}
with initial condition $T(0)=1$.
We conclude that,
\begin{equation}\label{51}
T= \frac{2\sinh\left(\frac{1}{2}\theta\sqrt{\Delta}\right)}{\theta\sqrt{\Delta}}.
\end{equation}
Note that the combination of the Weyl map with the obtained gauge operator is exactly the  Duflo quantization map for polynomial functions on the dual of $\mathfrak{su}(2)$  \cite{duflo}, extended to a larger class of functions  in \cite{FM} by means of Fourier expansion
\be
\chi_D(f)=\hat W\circ T(f)=\int \frac{d^{3}k}{\left( 2\pi
\right) ^{3}}\frac{2\sin\left(\frac{1}{2}\theta|k|\right)}{\theta|k|}\tilde{f}\left( k\right) e^{ik_{m}\hat{x}^{m}} \,.
\ee

The Duflo isomorphism \cite{duflo} establishes that  the algebra of invariant polynomials on the dual of finite dimensional  Lie algebras is  isomorphic to the center of the corresponding universal enveloping algebra. It is obtained on composing  the  Poincar\'e-Birkhoff-Witt isomorphism (which is only an isomorphism at the level of vector spaces)
with an automorphism of the space of polynomials. Let us shortly review the construction.
Let us indicate with $S(\mathfrak{g})$ the symmetric algebra over the Lie algebra
$\mathfrak{g}$. This may be identified with the polynomial functions on the dual
$S(\mathfrak{g})^*$. This is isomorphic to the universal enveloping algebra
$U(\mathfrak{g})$ as a vector space. Let us consider the subalgebra of
 $\mathrm{ad}_{\mathfrak{g}}$ invariant polynomials, $S(\mathfrak{g})^\mathfrak{g} $
and  let us define the Poincar\'e-Birkhoff-Witt map
\be
I_{PBW}: S(\mathfrak{g})^\mathfrak{g}\rightarrow \mathcal{Z}(U(\mathfrak{g}))
\ee
as the symmetrized product
\be
I_{PBW}(x_1,...x_n)= \frac{1}{n!}\sum_{\sigma(p)\in\Pi(n)} x_{\sigma(1)} ..x_{\sigma(n)}
\ee
with $\mathcal{Z}(U(\mathfrak{g}))$ the center of $U(\mathfrak{g})$.
This is an isomorphism of vector spaces but not in general an isomorphism of algebras.
Generalising previous results of Harish-Chandra to all finite dimensional Lie algebras
Duflo proved in \cite{duflo} that it could be extended to an algebra isomorphism.
Upon defining
\be
j^{\frac{1}{2}}(\tau\cdot\del):= {\det}^{\frac{1}{2}}\left[\frac{\sinh\frac{1}{2} \,
\tau\cdot \del}{\frac{1}{2} \, \tau\cdot \del}\right] \label{jDuf}
\ee
where
\be
\tau\cdot \del:= \tau_i\frac{\del}{\del x^i}
\ee
$\tau_i$ the generators of the Lie algebra in the adjoint representation  and $\del/\del x^i$ differential operators acting on the universal enveloping algebra,
the Duflo isomorphism
\be
\chi_D: S(\mathfrak{g})^{\mathfrak{g}}\rightarrow \mathcal{Z}( U({\mathfrak{g}}))
\label{duflo}
\ee
is proven to be
\be
\chi_D=I_{PBW}\circ j^{\frac{1}{2}}(\tau\cdot \del).\label{duflomap}
\ee
 Restricting to the
 $\mathfrak{su}(2)$ case, we can check by direct calculation that the Duflo automorphism $j^{\frac{1}{2}}(\tau\cdot\del)$ and the gauge operator $T$ given by Eq. \eqn{51} coincide.  Indeed,  the adjoint representation of  $\mathfrak{su}(2)$ is given by the defining representation of
 $\mathfrak{so}(3)$, that is
\be
( \tau_i)_{jk}= -i \epsilon_{ijk}. \,\;\;\;
\ee
On using
\be\label{trlog}
 {\det}^{\frac{1}{2}}\left[\frac{\sinh\frac{1}{2} \,
x}{\frac{1}{2} \, x}\right]=\exp\left[\frac{1}{2}\Tr\log \left(\frac{\sinh\frac{1}{2} \,
x}{\frac{1}{2} \, x}\right)\right]
\ee
and
\be
\Tr(\tau\cdot \del)^{2n}= 2\Delta^{n}
\ee
it is easily proven by series expansion that Eq. \eqn{jDuf} and Eq. \eqn{51} coincide order by order. Let us stress that, in order for the result to be correct, it is crucial that the representation of the Lie algebra generators be the adjoint one.

 This  result implies that
the closed star product we are looking for is exactly the one obtained from the Duflo quantization map,
\be
f\star g=\chi{_D}^{-1}\left(\chi_{D}(f)\cdot\chi_{D}(g)\right)
\ee
which, in principle, is an expected result. Since the integration measure $\mu(x)$ was chosen to be constant, the closed star product is the Kontsevich star product, as it was mentioned in the previous section. While, as is known, Duflo isomorphism has been demonstrated by  Kontsevich to follow from his formality theorem \cite{Kontsevich} when the Poisson manifold is the dual of a Lie algebra.

Taking into account the eqs. \eqn{w1} with \eqn{Bk} for the Weyl star product, the formula \eqn{D}, which relates the Weyl star product and the closed one, and the explicit form of the gauge operator $T$ given by \eqn{51} we may write the expression for the closed star product for $\mathfrak{su}(2)$:
\beqa \label{cstar}
f\star g(x)&=&\int \frac{d^{3}k_1}{\left( 2\pi
\right) ^{3}} \frac{d^{3}k_2}{\left( 2\pi
\right) ^{3}}\tilde{f}\left( k_1\right)\tilde{g}\left( k_2\right)\\
&\times&\frac{\sin\left(\theta|k_1|/2\right)\sin\left(\theta|k_2|/2\right)}{\theta|k_1||k_2|}\frac{2|B(k_1,k_2)|}{\sin\left(\theta|B(k_1,k_2)|/2\right)}e^{iB_m(k_1,k_2)x^m}.\nn
\eeqa
In particular,
\be\label{starexp}
e^{i\vec p\cdot \vec x}\star e^{i\vec q\cdot \vec x}=\frac{\sin\left(\theta|\vec p|/2\right)\sin\left(\theta|\vec q|/2\right)}{\theta|\vec p||\vec q|}\frac{2|\vec B(\vec p,\vec q)|}{\sin\left(\theta|\vec B(\vec p,\vec q)|/2\right)}e^{i\vec B(\vec p,\vec q)\cdot \vec x}
\ee
A similar expression for the Duflo star product of two plane waves was obtained in \cite{OR13} in the context of noncommutative Fourier transform for Lie groups.

One may also write the modified star product as
\beqa \label{starsu2}
f\star g &=& f\cdot g+\frac{i\theta}{2}\varepsilon^{ijk}x^k\partial_i f\partial_j g-\frac{\theta ^{2}}{8}x^ax^b\varepsilon ^{jla}\varepsilon ^{imb}\partial _{i}\partial
_{j}f\partial _{l}\partial _{m}g  \label{20} \\
& -&\frac{\theta^2}{12}x^a\varepsilon^{ila}\varepsilon ^{jkl}\left( \partial _{i}\partial _{j}f\partial _{k}g-\partial
_{i}\partial _{j}g\partial _{k}f\right) -\frac{\theta ^{2}}{12}\partial _{i}f\partial
_{i}g  \nn\\
&-&\frac{i\theta^3}{8}\left[\frac{1}{6} x^a\varepsilon^{nka}\varepsilon ^{jmn}\varepsilon
^{ilm}\left( \partial _{i}\partial _{j}f\partial _{k}\partial _{l}g-\partial
_{i}\partial _{j}g\partial _{k}\partial _{l}f\right) \right.  \nn \\
&+&\frac{1}{3}x^a\varepsilon ^{nka}
\partial _{l}\partial _{n}f\partial _{k}\partial _{l}g
\nn \\
&+&\frac{1}{3}x^ax^b\varepsilon ^{lna}\varepsilon ^{jml}\varepsilon ^{ikb}\left(
\partial _{i}\partial _{j}f\partial _{k}\partial _{n}\partial _{m}g-\partial
_{i}\partial _{j}g\partial _{k}\partial _{n}\partial _{m}f\right)
\nn \\
& +& \left. \frac{1}{6}x^ax^bx^c\varepsilon ^{jla}\varepsilon ^{imb}\varepsilon ^{knc}\partial _{i}\partial
_{j}\partial _{k}f\partial _{l}\partial _{n}\partial _{m}g  \right]+O\left(\theta^4\right)
~.  \nn
\eeqa
From this expression we obtain for quadratic polynomials
\begin{equation}\label{starx}
    x^i\star x^j=x^i\cdot x^j+\frac{i\theta}{2}\varepsilon^{ijk}x^k-\frac{\theta^2}{12}\delta^{ij}.
\end{equation}
In particular this result yields for the Casimir function:
\be
\sum_i x^i\star x^i=\sum_i x^i\cdot x^i-\frac{\theta^2}{4} \label{Duflosymb}
\ee
in agreement with the Duflo quantization of  the  $\mathfrak{su}(2)$ Casimir function as a polynomial in the symmetric algebra. On applying \eqn{duflomap} and expanding   \eqn{trlog}    up to second order (higher terms yield zero when acting on quadratic polynomials), we have indeed
\be
\chi_D(x^j x^k)= I_{PBW} \circ \left(1+\frac{\del_i\del_i}{24}\right)(x^j x^k)=  I_{PBW}(x^j x^k) +\frac{1}{12}\delta^{jk }
\ee
so that, when applied to the quadratic Casimir function $\sum _i x^i x^i$  it yields
\be
\chi_D(\sum _i x^i x^i) = \sum_i \hat X^i \hat X^i + \frac{1}{4} \mathbb{I}. \label{dufloq}
\ee
with  $\hat X^i$ the usual angular momentum operators and $\mathbb{I}$ the identity operator, both in the appropriate representation. Solving for the Casimir operator we can see that Eq. \eqn{Duflosymb} is precisely the symbol of Eq. \eqn{dufloq}.

Because of its very definition the Duflo map represents  a mathematically preferred quantization scheme. On the physics side it has been recently  applied in \cite{RV}  to the quantization of the hydrogen atom  on the basis of the $SO(4)$ symmetry of the latter and it has been shown to correctly reproduce the spectrum. It is therefore remarkable that the closed star product \eqn{starsu2}  be singled out as the one  in agreement with  such a quantization scheme.

For applications to field theory we derive   the expression for the ordinary Fourier transform of the star product \eqn{cstar}. We have
 \beqa \label{eq1}
\mathcal{F}[f\star g](s)&=&\int d^{3}x \left[f\star g\right](x) e^{-is_m x^m}
\\&=&\int  \frac{d^{3}k_1}{\left( 2\pi
\right) ^{3}} \frac{d^{3}k_2}{\left( 2\pi
\right) ^{3}}\tilde{f}\left( k_1\right)\tilde{g}\left( k_2\right)(2\pi)^3\delta\left(s_m-B_m(k_1,k_2)\right)\nn\\
&\times&\frac{\sin\left(\theta|k_1|/2\right)\sin\left(\theta|k_2|/2\right)}{\theta|k_1||k_2|}\frac{2|B(k_1,k_2)|}{\sin\left(\theta|B(k_1,k_2)|/2\right)}.\nn
\eeqa
Using the presence of delta function, $\delta(s_m-B_m(k_1,k_2))$, on may integrate over $k_2$. To this end one needs to solve the equation
\be\label{eq2}
s_m=B_m(k_1,k_2),
\ee
w.r.t. $k_2$ in terms of $k_1$ and $s$. First we rewrite \eqn{eq2} as
\be\label{eq3}
\frac{\sin\left(\frac{\theta}{2}|s|\right)}{|s|}\vec s=\sqrt{1-| p_1|^2} \vec p_2+\sqrt{1-| p_2|^2}\vec p_1-\vec p_1\times\vec p_2,
\ee
where
\be\label{eq4}
\vec p_a=\vec k_a\sin\left(\frac{\theta}{2}|k_a|\right)/|k_a|.
\ee
By defining $\vec c=\vec s \sin\left({\theta}|s|/{2}\right)/|s|$ we may solve \eqn{eq3} with respect to $p_2$,
\be\label{eq5}
\vec p_2=\vec c+ \vec p_1 \frac{\vec p_1\cdot\vec c-\sqrt{1-| p_2|^2}}{\sqrt{1-| p_1|^2}}-\vec p_1\times\vec c.
\ee
What is missing here is to find the expression for $e=\sqrt{1-| p_2|^2}$ in terms of $p_1$ and $c$. Considering the square of the both sides of \eqn{eq5} one gets the quadratic equation for $e$. Having the expression for $\vec p_2$ in terms of $\vec p_1$ and $\vec s$ and the equation \eqn{eq4} we finally write
\be\label{eq6}
\vec k_2=\vec k_2(k_1,s)=\left.\frac{2\arcsin| p_2|}{\theta| p_2|} \vec p_2\right|_{\vec p_1=\vec k_1\sin\left(\frac{\theta}{2}|k_1|\right)/|k_1|}.
\ee
The perturbative expression reads
\be
\vec k_2=\vec s- \vec k_1 +\frac{i\theta}{2} \vec k_1 \times\vec s+\dots
\ee

Substituting \eqn{eq6} in \eqn{eq1} we end up with
\be\label{eq7}
\mathcal{F}[f\star g](s)=\int  \frac{d^{3}k_1}{\left( 2\pi
\right) ^{3}}\tilde{f}\left( k_1\right)\tilde{g}\left( k_2\right)\frac{2|s|\sin\left(\theta|k_1|/2\right)\sin\left(\theta|k_2|/2\right)}{\theta|k_1||k_2|\sin\left(\theta|s|/2\right)},
\ee
where $\vec k_2$ is given by \eqn{eq6}.

\section{Derivative operator and deformed Leibniz rule}\label{derop}

To fix the derivative operator $\hat\partial_i$ on $\mathbb{R}^3_\theta$ we impose two natural requirements. First, it should be self-adjoint with respect to the introduced trace functional, i.e.,
\begin{equation}\label{21}
  \Tr\left(  \hat\partial_i f \star g\right)=\Tr\left(  f \star\hat\partial_i g\right).
\end{equation}
The second, is that it should have a correct commutative limit, in a sense that the algebra of commutation relations between the derivatives and coordinate operators in the limit $\theta \rightarrow0$, should reproduce the standard Heisenberg algebra. The operator satisfying these two requirements can be chosen as \be\hat\partial_{i}=-i\partial _{i}-\frac{i}{2}\partial _{i}\ln \mu \left( x\right) ,\ee see \cite{Kup14} for details. Taking into account the specific choice of the measure $\mu=1$ on $\mathbb{R}^3_\theta$, if one sets
\begin{equation}
\hat\partial_{i}=-i\partial _{i} ,
\label{p}
\end{equation}%
both conditions are satisfied.

Notice however  that ordinary  derivatives are not derivations of the star product \eqn{starsu2}, namely, they do not satisfy the standard Leibniz rule
 \begin{equation}\label{Leibniz}
    \partial_{i}\left(f\star g\right)\neq\left(\partial_{i}f\right)\star g+f\star\left(\partial_{i} g\right).
\end{equation}
We shall show  below that for closed star products ordinary derivatives can only violate the Leibniz rule  {\it weakly}, that is, the Leibniz rule is verified {\it up to a total derivative}.
This property has been already used in \cite{Kup17}  for the derivation of  conservation laws \cite{Kup17}. In next section, it will allow us to define the kinetic part of  a scalar field theory action as in the commutative case, in terms of the ordinary Laplacian.
The closure condition (\ref{3}), written as
\begin{equation}\label{22}
  \int\, \dd^3x\,\left[ f\star g-f\cdot g\right]=0,
\end{equation}
implies that the integrand is a total derivative, i.e., there exist such a bidifferential operator $a^j(f,g)$, that
\begin{equation}\label{trace2}
  f\star g=  f\cdot g+\partial_j a^j(f,g).
\end{equation}
Up to the third order in $\theta$ it can be written as
\begin{align}
& a^j(f,g)=\frac{i\theta}{4}x^l\varepsilon^{jkl}( f\partial_k g-\partial_k f g)+\frac{\theta^2}{16}x^ax^b\varepsilon ^{jka}\varepsilon ^{mnb}\left(\partial_k\partial_m f\partial_n g-\partial_m f\partial_k\partial_n g\right)\label{trace3}\\
&+\frac{\theta^2}{48}x^k\left(\varepsilon^{ilk}\varepsilon^{mjl}-\varepsilon^{mlk}\varepsilon^{jil}\right)\partial_m f\partial_i g+O\left( \theta ^{3}\right).\notag
\end{align}
Now differentiating  both sides of (\ref{trace2}) one finds:
\begin{equation}
 \partial_i[ f\star g]=\partial_i f\cdot g+ f\cdot\partial_i  g+\partial_j \partial_i a^j(f,g).\label{p1}
\end{equation}
Using (\ref{trace2}) one more time we end up with the identity
\begin{equation}
 (\partial_i f)\star g+ f\star ( \partial_i  g)=\partial_i[  f\star g]+\partial_j b^{ji} (f,g),  \label{p2}
\end{equation}
where
\begin{align}
&b^{ji}(f,g)=a^j\left(\partial_i f,g\right)+a^j\left( f,\partial_i g\right)-\partial_i a^j(f,g)\label{b}\\&=\frac{i\theta}{4}\varepsilon^{jki}( \partial_k f\cdot g -f\cdot\partial_k g)+O\left( \theta ^{2}\right).\notag
\end{align}
Notice that here the coefficients $a^j$ and $b^{ji}$ are given only at first nontrivial orders. Higher order contributions and the closed formula can be obtained from \eqn{22} taking into account the definition of the star product $\star$, \eqn{D}, the gauge operator $T$, given by \eqn{51}, and the Weyl star product for $\mathfrak{su}(2)$ in the form \eqn{40}, \eqn{42}.

\section{Scalar field theory on $\R^3_\theta$}\label{fieldtheo}
As an application
let us consider the classical action
\begin{equation}\label{71}
S=\int \dd^{3}x \left[\frac{1}{2}\partial_\rho \phi^*\star \partial^\rho \phi+\frac{m^2}{2}\phi^*\star \phi+\frac{\lambda}{4!}\phi^*\star \phi\star \phi^*\star \phi\right]
\end{equation}
with $\phi$ a scalar,  complex field.
Since the product is closed, we may integrate by parts and neglect total derivatives so to have, for the quadratic part
\be
S_K=\int \dd^{3}x \left[\frac{1}{2} \phi^*(\Delta +m^2) \phi\right]
\ee
As for the interaction term by the same reasons we can drop one star and write it as:
\be
S_{int}=\int \dd^{3}x \frac{\lambda}{4!}\left(\phi^*\star \phi\right)^2,
\ee
where the star product is given by \eqn{cstar}. Using \eqn{eq7} we may write it in terms of Fourier transform.

As an example of application of the deformed Leibniz rule \eqn{p2} lets us derive the electric charge density of the complex scalar field in the interacting theory \eqn{71}. The Euler-Lagrange equations for the scalar field $\phi$ and its conjugate $\phi^\ast$ read
\begin{eqnarray}\label{73}
  &&\Delta \phi-m^2 \phi-\frac{\lambda}{3!}\phi\star \phi^\ast\star \phi=0,\\
 && \Delta \phi^\ast-m^2 \phi^\ast-\frac{\lambda}{3!}\phi^\ast\star \phi\star \phi^\ast=0.\notag
\end{eqnarray}
Let us star multiply the first equation from the right by $\phi^\ast$ and then subtract the second equation star multiplied from the left by $\phi$. Taking into account the associativity of the star product we end up with the identity
\be\label{74}
\Delta \phi\star\phi^\ast-\phi\star\Delta \phi^\ast=0.
\ee
Using \eqn{p2}, the LHS of this equation can be represented as
\begin{eqnarray}\label{75}
&&\Delta \phi\star\phi^\ast+\partial_i\phi\star\partial^i\phi^\ast-\partial_i\phi\star\partial^i\phi^\ast-\phi\star\Delta \phi^\ast\\
&&=\partial_i\left(\partial_i\phi\star\phi^\ast-\phi\star\partial^i\phi^\ast+b^{ij}(\phi,\partial_j\phi^\ast)-b^{ij}(\partial_j\phi,\phi^\ast)\right).
   \notag
\end{eqnarray}
That is, \eqn{74} becomes the conservation law, $\partial_i j^i_{\theta}=0$, for the noncommutative current
\be\label{76}
j^i_{\theta}=\partial_i\phi\star\phi^\ast-\phi\star\partial^i\phi^\ast+b^{ij}(\phi,\partial_j\phi^\ast)-b^{ij}(\partial_j\phi,\phi^\ast).
\ee
In the commutative limit, $\theta \rightarrow0$, $j^i_{\theta}$ transforms to the standard current density for the complex scalar field, $j^i_{0}=\partial_i\phi\phi^\ast-\phi\partial^i\phi^\ast$. However, it should be noted that the commutative current density $j^i_{0}$ is no longer conserved in the interacting theory \eqn{71}.

\section{Conclusion and perspectives}

One of the main results of the article is represented by  the closed expression for the gauge operator $T$ \eqn{51} which relates the Weyl star product with the closed one, together with the fact that the closed star product is exactly the one we would obtain by Duflo de-quantization. Because of the existence of a closed star product, we could  define a derivative operator which obeys a deformed Leibniz rule. This  in turn allowed the definition of a Laplacian that is exactly the ordinary non-deformed one. In such a framework it becomes now possible to  study interacting quantum   field theories,  which are deformations of ordinary commutative ones,  the kinetic term of the action being undeformed. As an example one could consider the scalar action   introduced in section \ref{fieldtheo}. In the latter case the quartic  interaction term reduces to the square of $\phi^*\star\phi$ which can in turn be analyzed in terms of  the Fourier transform \eqn{eq7}.  Another possibility would be to look for a suitable matrix basis adapted to the quantization-dequantization scheme considered here (see \cite{LV14} for details).
We shall come back to this issue elsewhere.

\subsection*{Acknowledgements}

The work of V.G.K.\ was
supported by FAPESP and CNPq. P.V. acknowledges partial support by  Compagnia di San Paolo in the framework
of the program STAR 2013.


\begin{thebibliography}{99}

\bibitem{QGrav} L.~Freidel and E.~R.~Livine,
  {\it Effective 3-D quantum gravity and non-commutative quantum field theory,}
\textit{  Phys.\ Rev.\ Lett.} \  {\bf 96} (2006) 221301
  [hep-th/0512113].
  %%CITATION = HEP-TH/0512113;%%
  \\
  A.~Baratin and D.~Oriti,
  {\it Group field theory with non-commutative metric variables,
  Phys.\ Rev.\ Lett.}\  {\bf 105} (2010) 221302
  [arXiv:1002.4723 [hep-th]].
 %%CITATION = ARXIV:1002.4723;%%
 \bibitem{OR13} C.~Guedes, D.~Oriti and M.~Raasakka,
 {\it Quantization maps, algebra representation and non-commutative Fourier transform for Lie groups, \,
 J.\ Math.\ Phys.}\  {\bf 54} (2013) 083508.
 [arXiv:1301.7750 [math-ph]].
 %%CITATION = ARXIV:1301.7750;%%

 \bibitem{RV} L.Rosa and P.Vitale, {\it On the $\star$-product quantization and the Duflo map in three dimensions, Mod. Phys. Lett.} \textbf{A27} (2012) 1250207.
  %%CITATION = ARXIV:1209.2941;%%

\bibitem{GP13} V.~G\'alikov\'a, S.~Kov\'a$\check{c}$ik and P.~Pre$\check{s}$najder,
  {\it Laplace-Runge-Lenz vector in quantum mechanics in noncommutative space,
  J.\ Math.\ Phys.}\  {\bf 54} (2013) 122106
  [arXiv:1309.4614 [math-ph]].
  %%CITATION = ARXIV:1309.4614;%%

  \bibitem{Kup15} V.G. Kupriyanov, {\it Hydrogen atom on curved noncommutative space, J.Phys. A: Math. Theor.}  \textbf{46} (2013) 245303.

\bibitem{Vitale1} P. Vitale, J.-C. Wallet, {\it Noncommutative field theories on $\mathbb{R}^3_\lambda$. Towards UV/IR mixing freedom, JHEP} 1304 (2013) 115.
[arXiv:1212.5131 [hep-th]].
  %%CITATION = ARXIV:1212.5131;%%
  \\
P. Vitale,
  {\it Noncommutative field theory on $\mathbb{R}^3_\lambda$,
  Fortsch.\ Phys.} \  {\bf 62} (2014) 825
  [arXiv:1406.1372 [hep-th]].
  %%CITATION = ARXIV:1406.1372;%%


\bibitem{Vitale2} A. G\'er\'e, P. Vitale, J.-C. Wallet, {\it Quantum gauge theories on noncommutative 3-d space,
  Phys.\ Rev.} \  {\bf D 90} (2014) 045019
  [arXiv:1312.6145 [hep-th]].
  %%CITATION = ARXIV:1312.6145;%%

\bibitem{Jabari} A.B. Hammou, M. Lagraa, M.M. Sheikh-Jabbari, {\it Coherent state induced star product on $\R^3_\lambda$ and the fuzzy sphere, Phys.Rev.}  \textbf{D 66} (2002) 025025.

\bibitem{Felder} G. Felder, B. Shoikhet, {\it Deformation Quantization with Traces, Lett. Math. Phys.}  \textbf{53} (2000)
75-86 .

\bibitem{Kontsevich} M.~Kontsevich, {\it Deformation quantization of Poisson manifolds,
Lett.\ Math.\ Phys.} \ \textbf{66} (2003) 157 [arXiv:q-alg/9709040].

\bibitem{Dito} G. Dito, {\it Kontsevich star product on the dual of a Lie algebra, Lett. Math. Phys.}  {\bf 48} (1999) 307–322

\bibitem{Gutt} S. Gutt, {\it An explicit $\star$-product on the cotangent bundle of a Lie group, Lett. Math. Phys.} {\bf  7}  (1983) 249–258

\bibitem{Kup17} V.G. Kupriyanov, {\it Dirac equation on coordinate dependent noncommutative space-time, Phys.Lett.} {\bf B 732} (2014) 385.

\bibitem{KV} V.G. Kupriyanov, D.V. Vassilevich, {\it Star products made
(somewhat) easier,  Eur.Phys.J.}  \textbf{C 58} (2008) 627.

\bibitem{Meljanac}  N.~Durov, S.~Meljanac, A.~Samsarov and Z.~Skoda,
{\it A universal formula for representing Lie algebra generators as formal
power  series with coefficients in the Weyl algebra,  J.\ Algebra}  \textbf{%
309} (2007) 318

\bibitem{Meljanac1} S. Meljanac, A. Samsarov, M. Stojic, K.S. Gupta, {\it kappa-Minkowski space-time and the star product realizations, Eur.Phys.J.}  {\bf C 53} (2008) 295

\bibitem{GLMV} J.~M.~Gracia-Bondia, F.~Lizzi, G.~Marmo and P.~Vitale,
  {\it Infinitely many star products to play with,
  JHEP}  {\bf 0204} (2002) 026
  [hep-th/0112092].
  %%CITATION = HEP-TH/0112092;%%

\bibitem{Meljanac2} S.~Meljanac and M.~Stojic,  {\it New realizations of
Lie algebra kappa-deformed Euclidean space,  Eur.\ Phys.\ J.} \  \textbf{C 47}
(2006) 531.

\bibitem{Behr} W.Behr, A.Sykora, {\it Construction of gauge theories on curved noncommutative space-time,\, Nucl.Phys. } {\bf B 698} (2004) 473

\bibitem{FM}
  L.~Freidel and S.~Majid,
  {\it Noncommutative harmonic analysis, sampling theory and the Duflo map in 2+1 quantum gravity,
  Class.\ Quant.\  Grav.}  {\bf 25} (2008) 045006
  [hep-th/0601004].

\bibitem{Pinzul} A. Pinzul, A. Stern, {\it Gauge Theory of the Star Product, Nucl.Phys.} {\bf  B 791} (2008) 284

\bibitem{DS} G. Dito and D. Sternheimer, {\it Deformation quantization:
Genesis, developments and metamorphoses,}  9--54, IRMA Lect. Math. Theor.
Phys., 1, de Gruyter, Berlin, 2002. [arXiv:math/0201168].

\bibitem{duflo} M. Duflo,
{\it Caract\`eres des alg\`ebres de Lie r\'esolubles}, C. R. Acad. Sci. Paris  S\`er.  A-B 269
(1969), A437.
12. \\
{\it Op\'erateurs diff\'erentiels bi-invariants sur un groupe de Lie}, Ann. Sci. \'Ecole Norm.
Sup.  (4) 10 (1977), no. 2, 265-288.

\bibitem{Kup14}
  V.~G.~Kupriyanov, {\it Quantum mechanics with coordinate dependent noncommutativity,
  J.\ Math.\ Phys.}\  {\bf 54} (2013) 112105.

  \bibitem{LV14} F.~Lizzi and P.~Vitale,
  {\it Matrix Bases for Star Products: a Review,
  SIGMA}  {\bf 10}, 086 (2014)
  [arXiv:1403.0808 [hep-th]].
  %%CITATION = ARXIV:1403.0808;%%

\end{thebibliography}
\end{document}